# Geometrical Asymmetry Effect on Energy and Momentum Transfer Rates in a Double-quantum-well Structure: Linear Regime

T. Vazifehshenas [1], T. Salavati-fard [2]*

[1] *Department of Physics, Shahid Beheshti University, G. C., Evin, 1983969411, Tehran, Iran;*

[2] *Department of Physics and Astronomy, University of Delaware, Newark, DE 19716 USA;*

* Corresponding author: Tel: +1 302 8316381; Fax: +1 302 8316381; E-mail: *taha@udel.edu*

**Abstract**

We investigate theoretically the effect of spatial asymmetry on the energy and momentum transfer rates in a double-quantum-well system using balance equation approach. Our study is limited to the linear regime where the applied electric field is sufficiently weak. We calculate the screened potential by using the random phase approximation and Hubbard approximation for the cases of high and low electron densities, respectively. Our numerical results predict that the spatial asymmetry affects considerably both the energy transfer and drag rates as a result of changes in plasmon modes. Also, we find that the spatial asymmetry effect disappears at lower temperatures by inclusion the short-range interaction.

1. Introduction

It has been more than two decades that nanostructures play a significantand promising role in the realmof science and technology. Double-quantum-well (DQW)structure which is one the most interesting members of semiconductor nanostructures family, has being broadly studied during past years [1-17].This structure consists of two coupled parallel quantum layerswhich areseparated from each other by several nanometers.In this coupled system, the inter-layer interaction creates several outstandingmany-body effects such as momentum transfer or Coulomb drag and also energy transfer betweenadjacent layers.In coupled quantum systems, under certain circumstances, when an external electric field is being applied to one of them, surprisingly, a voltage difference can be measured in the other system due to momentum transfer phenomenon. There are many papers describing various aspects of momentum transferusing experimental and theoretical techniques [4-13]. It is quite interesting that different electronic temperatures in two layers yields to the electronic energy transfer effect between electrons in different layers,interacting with each other through Coulomb potential, located in close vicinity and withoutany possibilities for inter-layer tunneling. This is a hot electron transport effectand can be happened in linear and non-linear regimes, corresponding to the presence and absence of an applied electric field, respectively [4, 14]. Utilizing the energy-balance approach, the behavior of the energy transfer rate in double quantum systems such as DQW, double-quantum-wire and double-layer graphene has been investigated in a few papers [15-21].

In spite of the fact that the symmetric DQW structure has been vastly studied, surprisingly, there are limited studies on the many-body properties of spatially asymmetric DQW structure [22-24] and therefore it is still exciting to investigate these many-particle nanoscale effects in an asymmetric geometry.The importance of studying the geometrical asymmetric structures originates from the fact that most real systems are not completely symmetric.Also, beingslightly asymmetric rather than perfectly symmetric could be considered as an additionalflexibility for engineering the system.As a result, considering spatial asymmetry into theoretical calculations might lead to an insight about the real and more interesting systems.

Motivated by the above mentioned ideas, in the research presented here we study the Coulomb drag and energy transfer rate in a geometrically asymmetric DQW structure and compare our numerical results with calculations for the symmetric system [13, 20]. The goal is to conduct systematically theoretical study in the effect of geometrical asymmetry on the energy and momentum transfer rates. Here, the study of energy and momentum transfer rates is limited to linear regime with sufficiently weak external electric field and includes contributions of both plasmon and quasi-particle excitations through dynamic screening approximationsfor the dielectric matrix of the system. Also to show the significant contribution of plasmon excitations, we calculate energy transfer rate within static screening approximation, as well. Furthermore, we calculate the energy and momentum transfer rates beyond the RPA by including both the zero- and finite-temperature Hubbard local field correction factors,which take into account the effect of exchange short-range interactions in the screened potential.

The rest of the article is organized as following. In the next section, we describe the model andtheoretical formalism ofenergy transfer rate and Coulomb drag for an asymmetric DQW system. Also, we explain briefly the zero- and finite-temperature Hubbard local field corrections for a two-dimensional system which is known to be necessary at low electron densities. The section 3 is dedicated to provide numerical calculations for both asymmetric and symmetric systems at two different electron densities. Discussions are also included in section 3. Finally in section 4, the conclusion of our work is provided.

## 2. Theoretical formalism

The system we study can be modeled as two parallel n-type doped GaAs-based infinite square quantum wells of widths $L_1$ and $L_2$ which are separated by a distance $d$ along the $z$-axis. The two layers have the sheet electron densities $n_1$ and $n_2$ and the electron temperatures $T_1$ and $T_2$. We consider the case that only the lowest subband in each quantum well is occupied and assume two layers are placed close together to ensure an effective inter-layer interaction but far enough to prevent the electron tunneling between them. As mentioned earlier, the inter-layer Coulomb interaction in a DQW structure is responsible for the energy and momentum transfer phenomena. Considering the screening effect, the dynamic effective inter-layer interaction, $W_{12}(\mathbf{q},\omega)$, is defined as:

$$W_{12}(\mathbf{q},\omega) = \frac{V_{12}(q)}{\det[\varepsilon(\mathbf{q},\omega)]} \qquad (1)$$

where $\mathbf{q}$ is a two dimensional wave vector in the plane of quantum well, $V_{12}(q)$ is the unscreened inter-layer (off-diagonal) element of the Coulomb interaction matrix and $\varepsilon(\mathbf{q},\omega)$ is the dielectric matrix of the system. There are different many-body approximations for calculating the dielectric function of an interacting system. In the limit of high electron density, the random phase approximation (RPA) is an appropriate approach to consider the long-range correlation characteristic of an electron gas system. In a double layer system, the determinant of the RPA dielectric matrix is given by [2]:

$$\det[\varepsilon(\mathbf{q},\omega,T_1,T_2)] = (1 - V_{11}(q)\chi_1(\mathbf{q},\omega,T_1))(1 - V_{22}(q)\chi_2(\mathbf{q},\omega,T_2)) - V_{12}(q)\chi_1(\mathbf{q},\omega,T_1)V_{21}(q)\chi_2(\mathbf{q},\omega,T_2) \qquad (2)$$

Here $\chi(\mathbf{q},\omega,T)$ is the finite-temperature dynamic two dimensional Lindhard polarization function and the explicit forms of its real and imaginary parts can be found in literature [2]. The above matrix elements of the unscreened electron-electron interaction, $V_{ij}(q)$, are of the form:

$$V_{ij}(q) = \frac{2\pi e^2}{\kappa q} F_{ij}(q) \qquad (3)$$

where $i, j = 1, 2$ are the layer indices, $\kappa$ is the dielectric constant of the host semiconductor and $F_{ij}(q)$ represents the elements of form factor matrix which basically depend on the geometrical parameters of the system and defined as [2]:

$$F_{ij}(q) = \iint dz\,dz' |\zeta_i(z)|^2 |\zeta_j(z')|^2 \exp[-q(z-z')] \qquad (4)$$

In above equation $\zeta_i(z)$ is the envelope wave function of lowest subband in the $i$th layer. The analytic expressions for form factor functions have been derived only for a few spatially symmetric systems: for example, in a system of two infinite square wells of equal width, $L$, the diagonal and off-diagonal elements of the form factor matrix are obtained as [25]:

$$F_{ii}(x) = \frac{3x + 8\pi^2/x}{x^2 + 4\pi^2} - \frac{32\pi^4[1-\exp(-x)]}{x^2(x^2 + 4\pi^2)^2} \qquad (5)$$

$$F_{ij}(x) = \frac{64\pi^4 \sinh^2(x/2)}{x^2(x^2 + 4\pi^2)^2} \exp(-qd) \qquad (6)$$

where $x = qL$. In the presence of spatial asymmetry *i.e.* the case we are interested in, the inter-layer form factor is obtained numerically from Eq. (4).

In this work, we are focusing on the effect of spatial asymmetry on the rates of energy and momentum transfer in a DQW system. These two important transport processes are strongly dependent on the inter-layer screened Coulomb interactions. Using the balance equation transport theory [26], the expressions for the energy transfer rate (power transfer) and the momentum transfer rate (Coulomb drag) can be derived in terms of the electron drift velocity and electron temperature. In our coupled double layer system, we assume an electric current is driven through one of layers ($v_{d1} \neq 0$) while no current flows in the other ($v_{d2} = 0$) but an induced electric field is formed in second layer ($E_2 \neq 0$). When the drive bias voltage is low enough to be within the linear regime, it is reasonable to take $v_{d1} \to 0$ at the end of calculations. From the balance equation transport theory [19, 27], the rate of energy transfer is given by ($\hbar = 1$):

$$P_{12}(v_{d1} - v_{d2}) = -\sum_{\mathbf{q}} \int_{-\infty}^{+\infty} \frac{\omega\,d\omega}{\pi} |W_{12}(\mathbf{q},\omega)|^2$$
$$\times \left[ n_B\left(\frac{\omega}{K_B T_1}\right) - n_B\left(\frac{\omega - \omega_{12}}{K_B T_2}\right) \right] \mathrm{Im}\,\chi_1(\mathbf{q},\omega,T_1)\,\mathrm{Im}\,\chi_2(-\mathbf{q},\omega_{12}-\omega,T_2) \qquad (7)$$

and also the rate of momentum transfer is obtained as

$$f_{12}(v_{d1}-v_{d2}) = -\sum_{\mathbf{q}}\int_{-\infty}^{+\infty}\frac{q_x d\omega}{\pi}|W_{12}(\mathbf{q},\omega)|^2$$
$$\times\left[n_B\left(\frac{\omega}{K_B T_1}\right)-n_B\left(\frac{\omega-\omega_{12}}{K_B T_2}\right)\right]\operatorname{Im}\chi_1(\mathbf{q},\omega,T_1)\operatorname{Im}\chi_2(-\mathbf{q},\omega_{12}-\omega,T_2) \quad (8)$$

Here $n_B(x)=1/(\exp(x)-1)$ is the Bose-Einstein distribution function and $\omega_{12}=q_x(v_{d1}-v_{d2})$. In the linear regime (weak electric field regime), the drag rate, $\tau_D^{-1}=-f_{12}(v_{d1})/(m_1 n_2 v_{d1})$, can be expressed by the following simple form [2]:

$$\tau_D^{-1} = -\frac{1}{8\pi^2 e^2 n_1 n_2 K_B T}\int_0^\infty q^3 dq \int_0^\infty d\omega \frac{|W_{12}(\mathbf{q},\omega)|^2}{\sinh^2(\omega/2K_B T)}\operatorname{Im}\chi_1(\mathbf{q},\omega,T)\operatorname{Im}\chi_2(\mathbf{q},\omega,T)$$
$$(9)$$

In the above equation, it is assumed that $T_1=T_2=T$.

It is well known that the exchange and correlation interactions are non-negligible at low electron density systems and therefore the RPA calculations ignoring short-range interactions is not reliable anymore. In this case, we can improve the RPA dielectric matrix by including the exchange effect through the Hubbard local field correction factor for a two dimensional electron gas system [29]:

$$G_H(q) = \frac{1}{2}\frac{q}{\sqrt{q^2+k_F^2}} \quad (10)$$

For this improvement one may replace the diagonal elements of interaction matrix, $V_{ii}$, by $V_{ii}(1-G_H)$ and leave the off-diagonal ones unchanged. By introducing a finite-temperature Fermi wave vector $k_F(T)=\sqrt{2m^*\mu(T)}$, the temperature dependent version of Hubbard approximation has been also obtained as follows [30]:

$$G_H(q,T) = \frac{1}{2}\frac{q}{\sqrt{q^2+2m^*\mu(T)}} \quad (11)$$

Here $\mu(T)$ is the chemical potential. It has turned out that accounting for the thermal effect in the Hubbard local filed correction, changes predictions by the zero temperature Hubbard approximation [20, 30-33].

### 3. Numerical results and discussion

We model our system as a spatially asymmetric DQW structure which consists of two parallel n-type doped GaAs-based quantum layers with equal electron densities that are coupled through the screened inter-layer Coulomb interaction. It is assumed that the square quantum wells are infinity deep so that we can safely ignore the electron tunneling between layers. To study an electron gas system, it is usual to work with the dimensionless density parameter, $r_s$, instead of electron density, $n$. $r_s$ is defined as the ratio of average distance between electrons in an electron gas without interaction to the effective Bohr radius $a_B^*$. The two dimensional version of this parameter has been calculated as $r_s = 1/(a_B^* \sqrt{\pi n})$. We choose two density parameters $r_s = 1$ and $r_s = 2$ for both layers and an inter-layer separation $d = 50$ nm which is kept fixed in our calculations. In case of $r_s = 1$ (*i.e.* high electron density), the RPA is a reliable approximation for calculating the dielectric matrix of a DQW system. However, for $r_s = 2$, it is necessary to include the short-range effect through the Hubbard local field correction. To see the effect of geometric asymmetry, we calculate the energy transfer and Coulomb drag rates for a few different structures denoted by $(L_1, L_2)$.

The energy transfer rate as a function of $T_2/T_F$, where $T_F$ is the Fermi temperature, for three different structures $(L_1, L_2) = (10\text{nm}, 50\text{nm})$, $(20\text{nm}, 40\text{nm})$ and $(30\text{nm}, 30\text{nm})$ and with an electron density $r_s = 1$ is illustrated in Fig. 1. The electron temperature of layer 1 is kept at $T_F$ and the dielectric matrix is calculated within the static RPA, thus the contribution of plasmon excitations is neglected. As it is shown, $P_{12}$ increases by increasing spatial asymmetry. The effect is stronger at low temperatures and becomes weaker by increasing temperature.

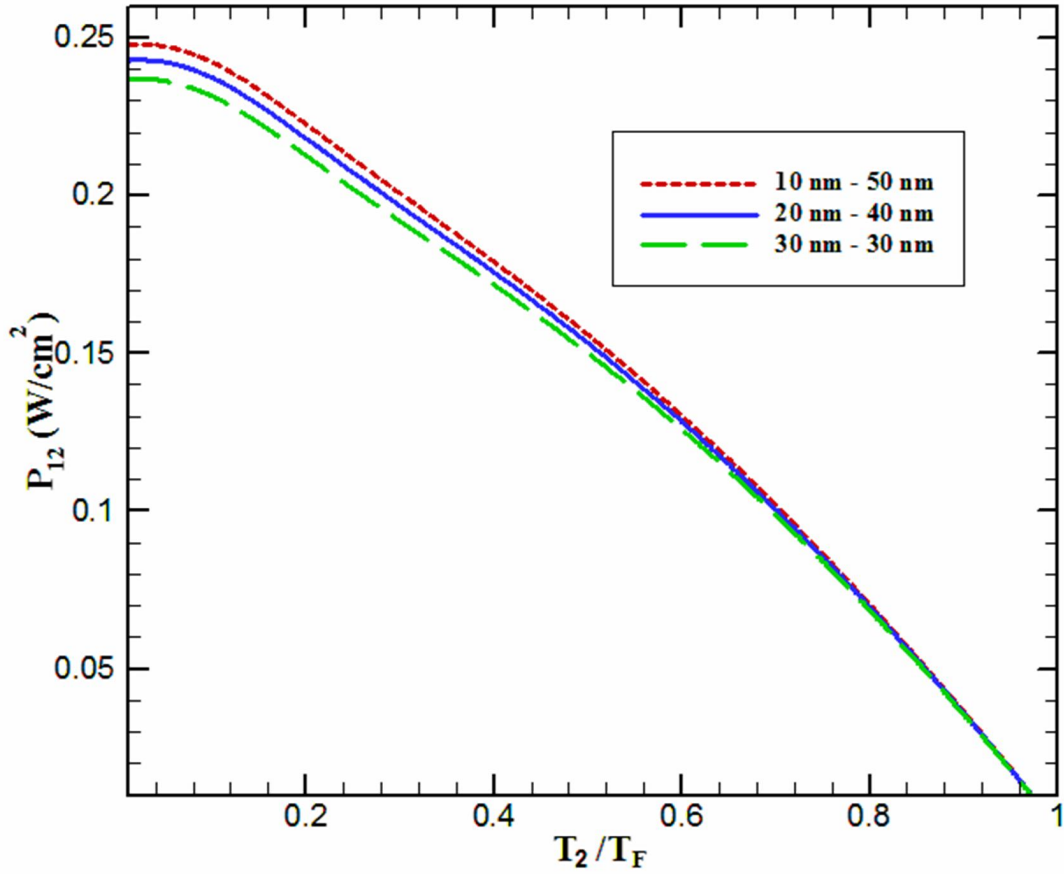

Fig.1. Energy transfer rate in units of $Watt/cm^2$ as a function of dimensionless electron temperature in layer 2 for three different structures $(L_1, L_2) = (10nm, 50nm)$, $(20nm, 40nm)$ and $(30nm, 30nm)$. The screened potential is calculated within the static RPA and electron temperature of layer 1 is fixed at $T_1 = T_F$ with $r_s = 1$ and also $d = 50$ nm.

Fig. 2 depicts the results of energy transfer rate calculation for three different structures $(L_1, L_2) = (10nm, 50nm)$, $(20nm, 40nm)$ and $(30nm, 30nm)$ at electron density $r_s = 1$. We have employed the dynamic RPA dielectric matrix to calculate screened interaction. At very low temperatures, $P_{12}$ increases by increasing space-asymmetry similar to what is happening in Fig. 1, however it changes its behavior approximately at $T_2 \approx 0.2 T_F$ (Fermi temperature for this electron density is around 125 K), so that the dynamic energy transfer rate decreases gradually by increasing the spatial asymmetry. The effect of asymmetry is really considerable in the interval $0.3 \leq T_2/T_F \leq 0.85$. This different behavior of $P_{12}$ within the dynamical RPA is associated with the plasmon contribution which is absent in the static approximation. The plasmons which are the collective excitations of a many-electron system and obtained from zeros of the dynamic dielectric function have a dominate effect on the transport properties of the system especially at long wavelengths. In a two component electron gas system such as DQW, the plasmon spectrum consists of two branches which usually called optical and acoustic plasmon modes corresponding to the in-phase and out-of-phase electron oscillations in two wells, respectively. As a result, considering both optical and acoustic plasmon modes contributions leads to an increase in the energy transfer rate up to an order of magnitude [19] and results in a large enhancement of the drag rate for $T_2 \geq 0.2T_F$ [2]. In the other hand, it is already known that geometric asymmetry has a significant influence on the plasmon modes in DQW structures [22-24]. In an asymmetric DQW, due to lack of inversion symmetry, the acoustic plasmon mode is much more affected than the optical mode [24]. If fact, the acoustic branch enters the Landau damping or single particle excitation region at a smaller wave vector compared to the acoustic plasmon curve of the symmetric system and therefore its contribution to the energy transfer rate decreases. This spatial asymmetry effect does not show up at temperatures below $0.2T_F$, because the electrons do not have enough energy to be excited very far above the Fermi surface. In addition, since at high temperatures ($T > 0.6T_F$) the plasmon modes are heavily damped due to Landau damping, their effect is weakened and hence make a negligible contribution to the energy transfer rate for temperatures above $0.85T_F$. The above discussion very well explains how the energy transfer rate is affected by the space asymmetry in the whole temperature interval.

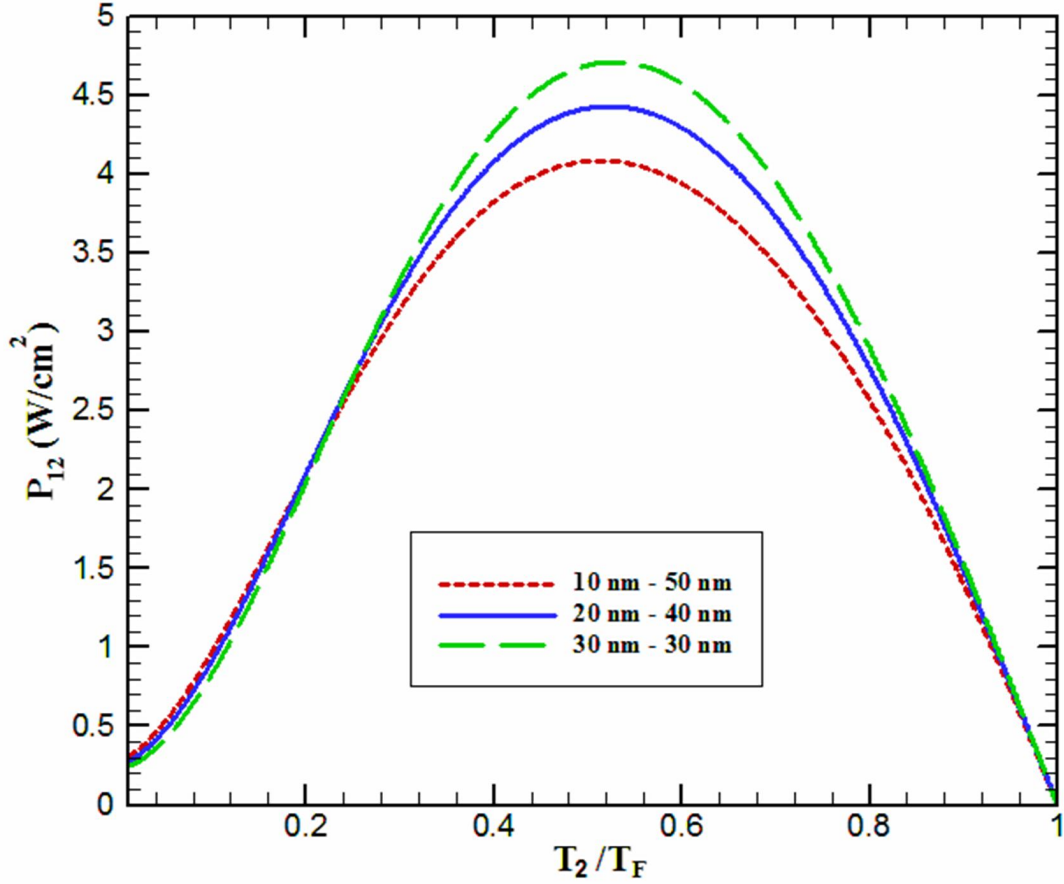

Fig. 2. Energy transfer rate in units of $\text{Watt}/\text{cm}^2$ as a function of dimensionless electron temperature in layer 2 for three different structures $(L_1, L_2) = (10\text{nm}, 50\text{nm})$, $(20\text{nm}, 40\text{nm})$ and $(30\text{nm}, 30\text{nm})$. The screened potential is calculated within the dynamic RPA and electron temperature of layer 1 is fixed at $T_1 = T_F$ with $r_s = 1$ and also $d = 50\,\text{nm}$.

In Fig. 3, the numerical calculations for $T^2$ − scaled Coulomb drag rate as a function of dimensionless temperature are illustrated. In drag rate calculations, it is assumed that $T_1 = T_2 = T$. Same as Figs. 1 and 2, calculations are performed within the RPA for three different structures with gradually increasing spatial asymmetry and at electron density $r_s = 1$. Here again, we include the contribution of both acoustic and optical plasmon excitations through the dynamic screening approximation. It is observed that the Coulomb drag rate decreases clearly by increasing the spatial asymmetry for temperatures higher than

$T_2 \approx 0.2T_F$. It is also seen that the effect of asymmetry disappears at high temperature similar to the previous figure. The same argument as what we have provided for energy transfer rate can be utilized to explain this behavior. In addition, Fig. 3 shows that the peak of drag rate shifts slightly to the higher temperatures by increasing the spatial asymmetry.

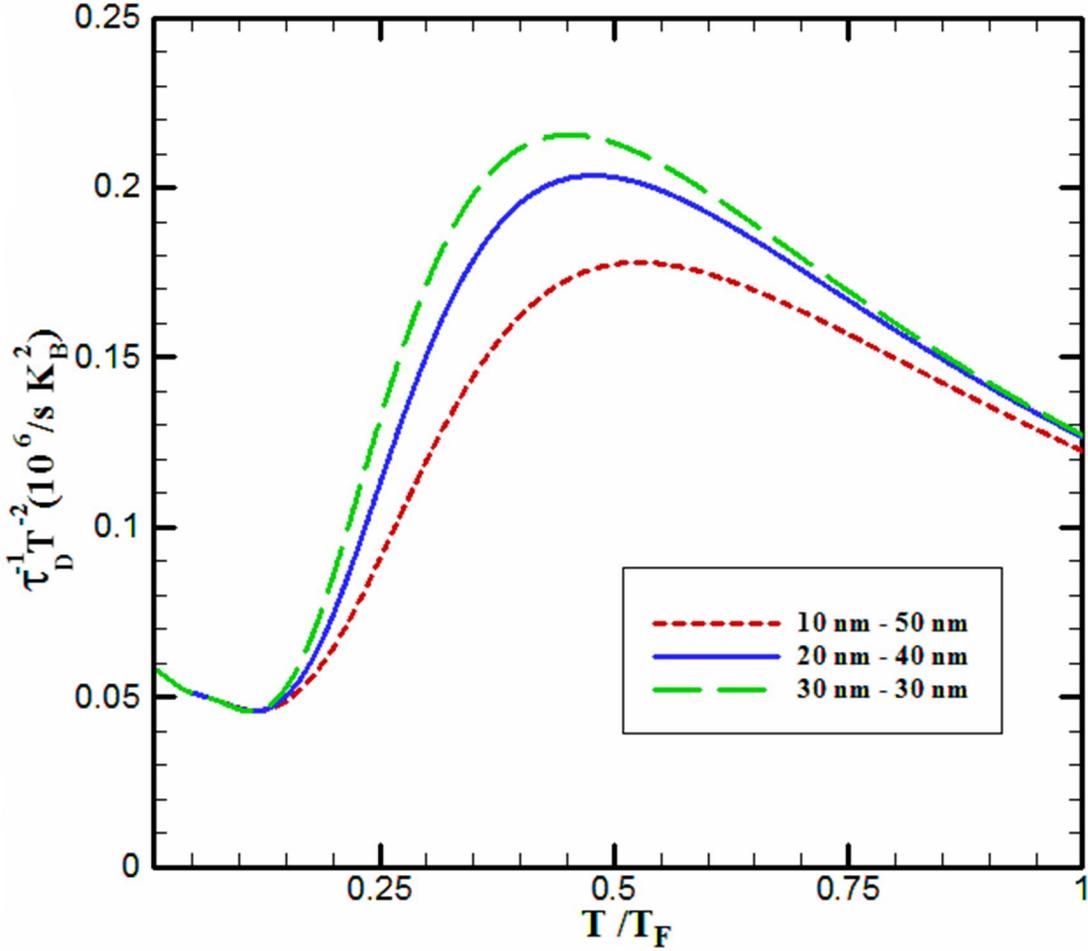

Fig.3. $\tau_D^{-1} T^{-2}$ as a function of electron temperature in units of $(10^6/sK_B^2)$ with $r_s = 1$ and $d = 50$ nm for three different structures $(L_1, L_2) = (10\text{nm}, 50\text{nm})$, $(20\text{nm}, 40\text{nm})$ and $(30\text{nm}, 30\text{nm})$. The screened potential is calculated within the dynamic RPA.

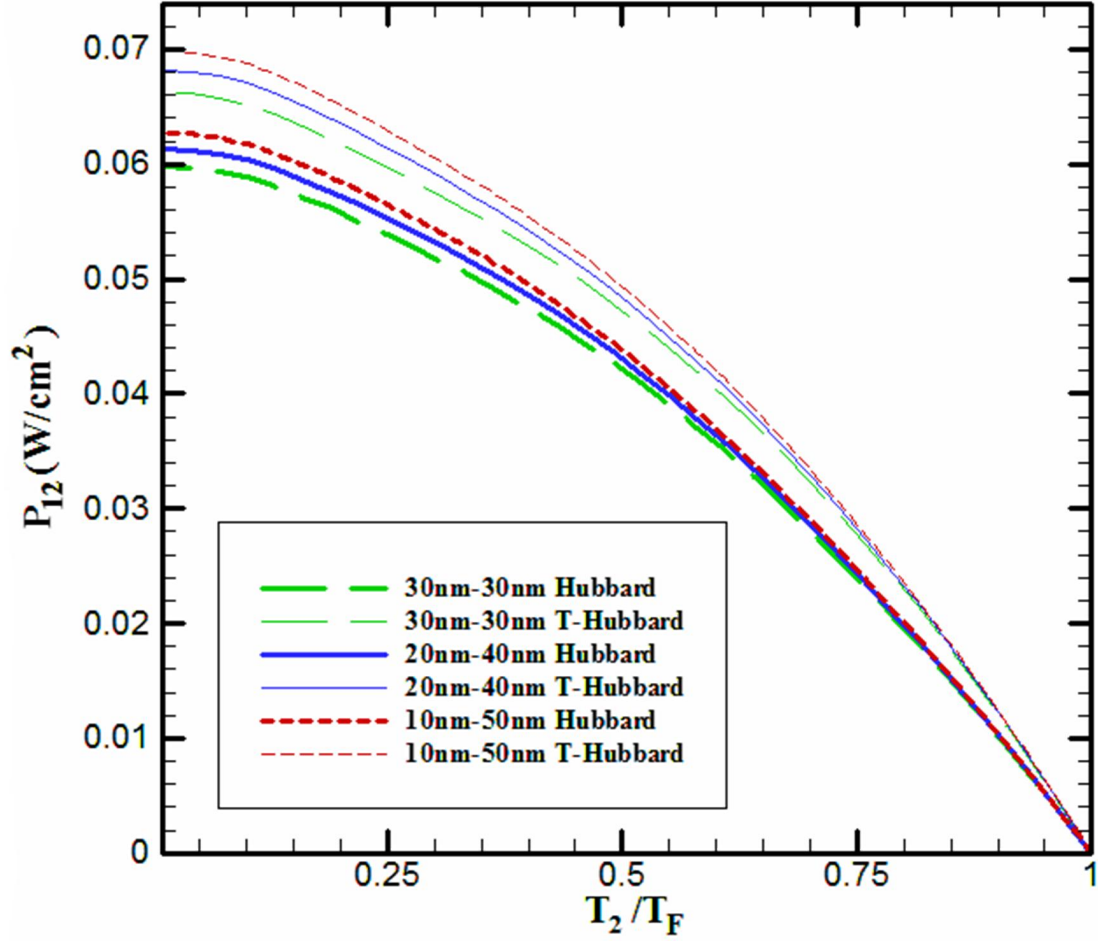

Fig. 4. Energy transfer rate in units of Watt/cm$^2$ as a function of dimensionless electron temperature in layer 2 for three different structures $(L_1, L_2) = $ (10nm,50nm), (20nm,40nm) and (30nm,30nm). $T_1 = T_F$, $r_s = 2$ and $d = 50$ nm. The statically screened potential is calculated within zero-temperature (thick curves) and finite-temperature (thin curves) Hubbard approximations.

To study the spatial asymmetry effect on the energy transfer and drag rates in a DQW system at lower electron densities, we change the density parameter to $r_s = 2$. As mentioned earlier, RPA is not reliable for this electron density and the short-range interactions should be taken into account; so we use the Hubbard-type of dielectric matrix.

Fig. 4 shows numerical calculations for $P_{12}$ within zero- and finite-temperature Hubbard approximations with static screened potential for three different structures. We can see the similar behavior of the static energy transfer rates obtained from both types of the Hubbard approximations to that from the RPA.

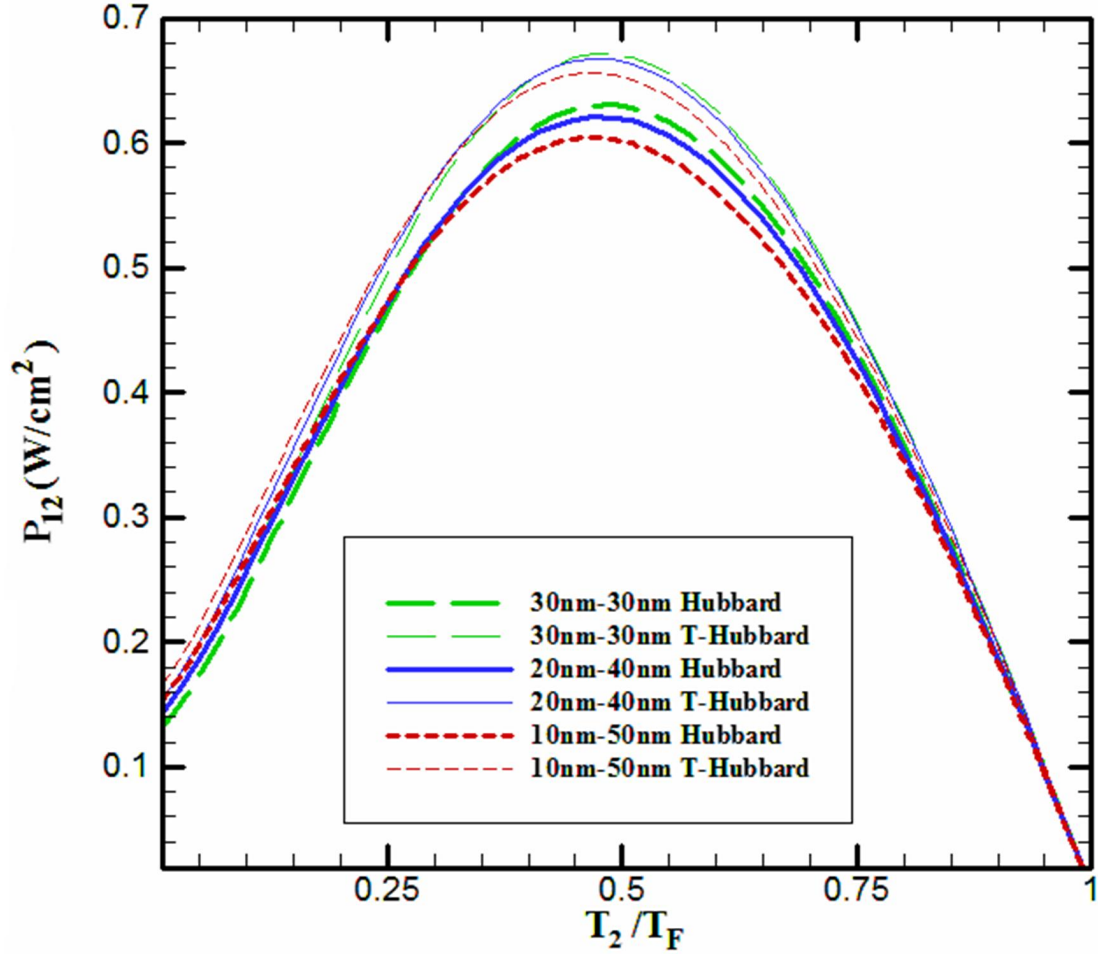

Fig. 5. Energy transfer rate in units of Watt/cm$^2$ as a function of dimensionless electron temperature in layer 2 for three different structures $(L_1, L_2) = (10\text{nm},50\text{nm})$, $(20\text{nm},40\text{nm})$ and $(30\text{nm},30\text{nm})$. $T_1 = T_F$, $r_s = 2$ and $d = 50$ nm. The dynamically screened potential is calculated within zero-temperature (thick curves) and finite-temperature (thin curves) Hubbard approximations.

We calculate the energy transfer rates for three different structures within both zero- and finite-temperature Hubbard approximations by employing the dynamic screened potential and show the results in Fig. 5. As reported previously for symmetric DQW [20], the energy transfer rate values obtained from the Hubbard approximation are greater than the RPA due to the stronger screening effect. Also, it is known that the values of the energy transfer rate in the finite-temperature Hubbard approximation are interestingly larger than in the zero-temperature one. The numerical calculations presented in Figs. 4 and 5 confirm that this trend is same for an asymmetric DQW as well. Taking a look at Fig. 5, it is clear that the behavior is similar to the RPA (Fig. 2). However, at this electron density ($r_s = 2$), the temperature at which the plasmons begin to contribute to the energy transfer rate is shifted to around $0.3T_F$ and $0.4T_F$ ($T_F \approx 31K$) with zero- and finite-temperature Hubbard local field corrections, respectively. It can be explained by this fact that lowering the electron density alters the plasmon dispersion relation and makes an increase in the plasmon energies so that their contributions starts at higher temperatures with respect to $r_s = 1$. Another interesting feature of calculations shown in Fig. 5 is that asymmetry effect on the dynamic energy transfer rate becomes weaker by increasing electron density parameter.

Finally, Fig. 6 depicts $T^2$– scaled Coulomb drag rate for two different structures $(L_1, L_2) = (10\text{nm}, 50\text{nm})$ and $(30\text{nm}, 30\text{nm})$ within the RPA, zero- and finite-temperature Hubbard approximations as a function of temperature. Calculations show that by going beyond the RPA, $T^2$– scaled Coulomb drag rate gets higher as was pointed out earlier [34]. Similar to Fig. 5 for dynamic energy transfer rate, it is clear that considering the local field corrections in our calculations weakens the effect of spatial asymmetry on the momentum transfer rate. It is also notable that the trend of asymmetry effect behavior remains the same in zero- and finite-temperature Hubbard approximations. According to this figure, the temperature at which the spatial asymmetry effect on the $T^2$– scaled Coulomb drag rate begins to show up is shifted toward higher temperatures by increasing the density parameter from $r_s = 1$ to $r_s = 2$ similar to energy transfer rate in Fig. 5. As the final remarkable point, it is found that by incorporating the Hubbard approximation, the effect of asymmetry on the energy transfer and drag rates disappears at lower temperatures because the short-range interactions push the acoustic plasmon mode into the single particle excitation region, resulting in Landau damping of the mode at smaller wavevectors compared to the RPA.

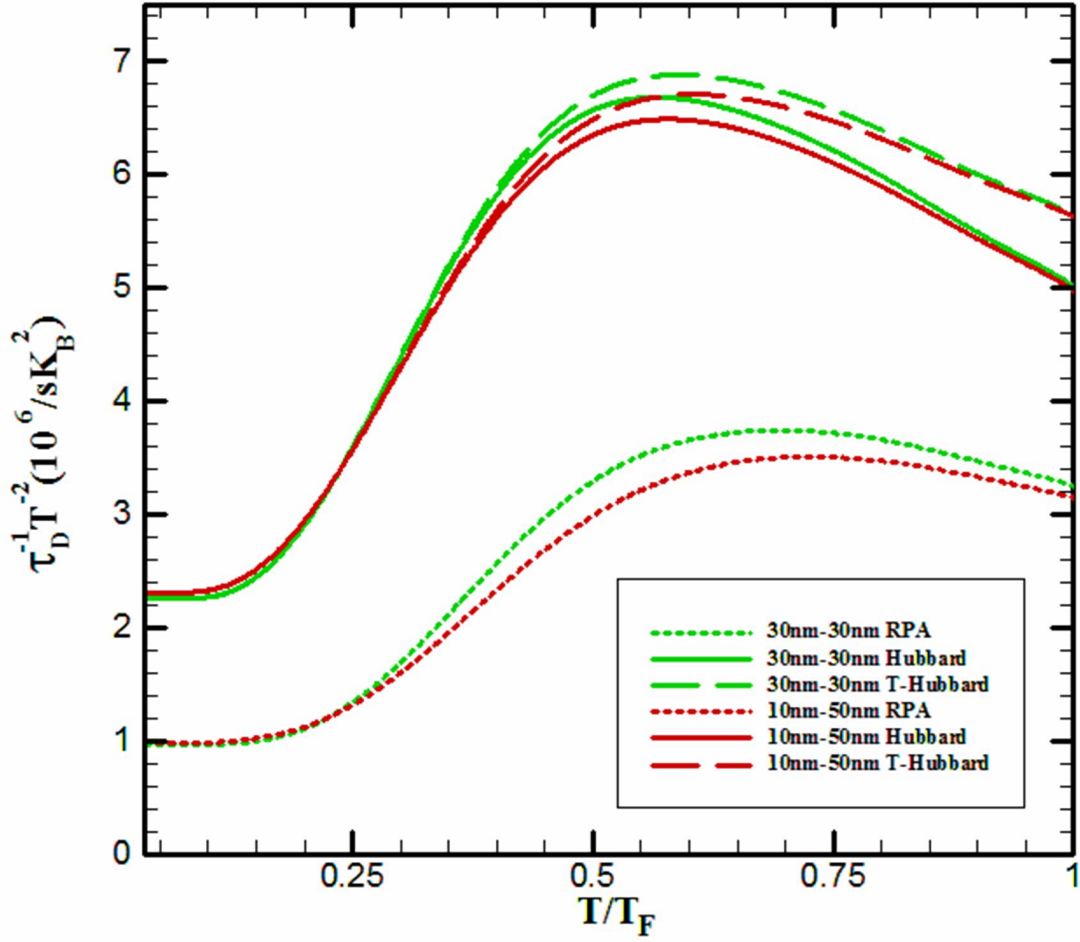

Fig. 6. $\tau_D^{-1}T^{-2}$ as a function of dimensionless electron temperature in units of $(10^6/sK_B^2)$ with $r_s = 2$ and $d = 50$ nm and for two different structures $(L_1, L_2) = (10\text{nm}, 50\text{nm})$ and $(30\text{nm}, 30\text{nm})$. The screened potential is calculated within the dynamic RPA, Hubbard and temperature-dependent Hubbard approximations.

## 4. Conclusion

In conclusion, we have studied theoretically the effect of spatial asymmetry on the energy and momentum transfer rates between two parallel coupled quantum layers in a drag experiment type setup. We have employed the energy and momentum balance equations for electrons and concentrated our study on the linear regime, where the applied electric field is sufficiently weak. To include the many-body effects on the effective electron-electron interactions, we have calculated the dielectric matrix within the RPA, zero-and finite-temperature Hubbard approximations. Our results predict that the geometric asymmetry affects significantly the energy transfer rate in both static and dynamic approximations and also the Coulomb drag rate. The energy transfer rate in static approximation increases notably by increasing geometric asymmetry at low temperatures and then the effect becomes weaker athigher temperatures. The changes in the energy transfer rate in the dynamic approximation and the Coulomb drag rate due to the asymmetry effect is different. They increase slightly by increasing spatial asymmetry at low temperatures and then change the behavior and decrease considerably at intermediate temperatures. We believe the contributions of plasmon excitations are responsible for this discrepancy in the static and dynamic cases. Furthermore, in all cases, the spatial asymmetry effect gradually disappears at high temperatures. And finally, it is obtained that by going beyond the RPA, the spatial asymmetry effect vanishes at lower temperatures.

**References**

[1] A. G. Rojo, J. Phys.: Condens. Matter **11,** R31 (1999).

[2] K. Flensberg and B. Y. Hu, Phys. Rev. B **52,** 14796 (1995).

[3] N. P. R. Hill, J. T. Nichols, E. H. Linfeld , K. M. Brown, M. Pepper, D. A. Ritchie, G. A. C. Jones, B. Y. Hu and K. Flensberg, Phys. Rev. Lett. **78,** 2204 (1997).

[4] P. M. Solomon, P. J. Price, D. J. Frank and D. C. La Tulipe, Phys. Rev. Lett. **63,** 2508 (1989).

[5] T. J. Gramila, J. P. Eisenstein, A. H. MacDonald, L. N. Pfeiffe and K. W. West, Phys. Rev. Lett. **66,**1216 (1991).

[6] U. Sivan, P. M. Solomon and H. Shtrikman, Phys. Rev. Lett. **68,** 1196 (1992).

[7] H. Noh, S. Zelakiewics, X. G. Feng, T. J. Gramila, L. N. Pfeiffer and K. W. West, Phys. Rev. B **58,** 12621(1998).

[8] H. C. TSo, P. Vasilopoulos and F. M. Peeters, Phys. Rev. Lett. **68,** 2516 (1992).

[9] A. Jauho and H. Smith, Phys. Rev. B **47,** 4420 (1993).

[10] L. Zhengand A. H. MacDonald, Phys. Rev. B **48,** 8203 (1993).

[11] A. Kamenev and Y. Oreg Phys. Rev. B **52,** 7516 (1995).

[12] L. Zheng and A. H. MacDonald,Phys. Rev. B **49,** 5522 (1994).

[13] T. Vazifehshenas and A. Eskourchi, Physica E **36,** 147 (2007).

[14] P. M. Solomon and B. Laikhtman, Superlattices Microstruct. **10,** 89 (1991).

[15] C. Jacobini and P. J. Price, Solid State Electron. **31,** 649 (1988).

[16] I. I. Boiko and Y. M. Sirenko, Phys. Stat. Sol. B **159,** 805 (1990).

[17] B. Laikhtman and P. M. Solomon, Phys. Rev. B **41,** 9921 (1990).

[18] B. Tanatar, J. App. Phys. **81** 6214 (1997).

[19] R. T. Senger and B. Tanatar, Solid State Commun. **121** 61 (2002).

[20] T. Vazifehshenas, B. Bahrami and T. Salavati-fard, Physica B **407**, 4611 (2012).

[21] B. Bahrami and T. Vazifehshenas, PhysLett A **376**, 3518 (2012).

[22] X. Liu, X. Wang and B. Gu, Eur. Phys. J. B **24**, 37 (2001).

[23] X. Liu, X. Wang and B. Gu, Eur. Phys. J. B **30**, 339 (2002).

[24] M. R. S. Tavares, G. Q. Haiand S. Das Sarma, Phys. Rev. B **64,** 045325 (2001).

[25] R. Jalabertand S. Das Sarma, Phys. Rev. B **40,** 9723 (1989).

[26] X. L. Lei and C. S. Ting, Phys. Rev. B **30,** 4809 (1984).

[27] C. S. Ting, *Physics of Hot Electron Transport in Semiconductors* (Singapore, World Scientific, 1992).

[28] J. Hubbard, Proc. R. Soc. London A **243,** 336 (1958).

[29] M. Jonson, J. Phys. C **9,** 3055 (1976).

[30] E. H. Hwang and S. Das Sarma, Phys. Rev. B **64,** 165409 (2001).

[31] A. Yurtsever, V. Moldoveanu and B. Tanatar, Phys. Rev. B **67,** 115308 (2003).

[32] T. Vazifehshenas and T. Salavati-fard, Phys. Scr. **81,** 025701 (2010).

[33] T. Vazifehshenas, Phys. Stat. Sol. (a) **205,** 1302 (2008).

[34] N. P. R. Hill, J. T. Nicholls, E. H. Linfield, M. Pepper, D. A. Ritchie, G. A. C. Jones, B. Yu-Huang-Hu and K. Flensberg, Phys. Rev. Lett. **78**, 2204 (1997).